\documentclass[iop]{emulateapj}

\usepackage{epsfig}

\def\simgt{\lower.5ex\hbox{$\; \buildrel > \over \sim \;$}}
\def\simlt{\lower.5ex\hbox{$\; \buildrel < \over \sim \;$}}

\def\etal{{et~al.}}
\def\amin{\ifmmode^{\prime}\else$^{\prime}$\fi}
\def\asec{\ifmmode^{\prime\prime}\else$^{\prime\prime}$\fi}

\def\simgt{\lower.5ex\hbox{$\; \buildrel > \over \sim \;$}}
\def\simlt{\lower.5ex\hbox{$\; \buildrel < \over \sim \;$}}

\newcommand\chandra{{\it Chandra}}

\newcommand\xmm{{\it XMM-Newton}}

\newcommand\nustar{{\it NuSTAR\/}}

\newcommand\hess{HESS}

\newcommand\fermi{{\it Fermi\/}}

\def\ztestone{59}
\def\ztesttwo{79}
\def\epochzero{56,466.0}
 \def\epochone{56,465.91194958}
 \def\epochtwo{56,564.29199072}

\def\freqzero{4.843950957(40)~Hz}

\def\perzero{0.206443048(33)~s}  
 \def\perone{0.206443040(33)}  
 \def\pertwo{0.206451335(17)}

\def\pdotv{$9.758(44) \times 10^{-13}$}
\def\fdotv{$-2.290(10) \times 10^{-11}$ Hz~s$^{-1}$}
\def\edotv{$4.4 \times 10^{36}$ erg~s$^{-1}$}
\def\bsv{$1.4\times 10^{13}$~G}
\def\taucv{3350~yr}

\def\period{206~ms}
\def\pdot{$\dot P =$ \pdotv}
\def\edot{$\dot E =$ \edotv}
\def\bs{$B_s =$ \bsv}
\def\tauc{$\tau_c \equiv P/2\dot P =$ \taucv}

\def\tev{HESS~J1640$-$465}

\def\cxo{CXOU~J164043.5$-$463135}
\def\fglone{1FGL~J1640.8$-$4634}
\def\fgltwo{2FGL~J1640.5$-$4633}
\def\fhl{1FHL~J1640.5$-$4634}
\def\snr{G338.3$-$0.0}
\def\psr{PSR~J1640$-$4631}

\slugcomment{Version of 2014 April 14}

\shorttitle{Discovery of a Pulsar in \tev}
\shortauthors{Gotthelf et al.}

\begin{document}

\title{NuSTAR Discovery of a Young, Energetic Pulsar Associated with \\ the 
Luminous Gamma-ray Source \tev}


\author{
E. V. Gotthelf\altaffilmark{1},
J. A. Tomsick\altaffilmark{2}, 
J. P. Halpern\altaffilmark{1}, 
J. D. Gelfand\altaffilmark{3,4},
F. A. Harrison\altaffilmark{5},
S. E. Boggs\altaffilmark{2},
F. E. Christensen\altaffilmark{6}, 
W. W. Craig\altaffilmark{2,7}, 
J. C. Hailey\altaffilmark{1}, 
V. M. Kaspi\altaffilmark{8}, 
D. K. Stern\altaffilmark{9}, 
W. W. Zhang\altaffilmark{10}
}
\altaffiltext{1}{Columbia Astrophysics Laboratory, Columbia University, 550 West 120th Street, New York, NY 10027-6601, USA; eric@astro.columbia.edu}
\altaffiltext{2}{Space Sciences Laboratory, University of California, Berkeley, CA 94720, USA}
\altaffiltext{3}{NYU Abu Dhabi, PO Box 129188, Abu Dhabi, UAE}
\altaffiltext{4}{Affiliate Member, Center for Cosmology and Particle Physics, New York University}
\altaffiltext{5}{Cahill Center for Astronomy and Astrophysics, California Institute of Technology, Pasadena, CA 91125, USA}
\altaffiltext{6}{DTU Space-National Space Institute, Technical University of Denmark, Elektrovej 327, 2800 Lyngby, Denmark}
\altaffiltext{7}{Lawrence Livermore National Laboratory, Livermore, CA 94550, USA}
\altaffiltext{8}{Department of Physics, McGill University, Montreal, QC H3A 2T8, Canada}
\altaffiltext{9}{Jet Propulsion Laboratory, California Institute of Technology, 4800 Oak Grove Drive, Pasadena, CA 91109, USA}
\altaffiltext{10}{NASA Goddard Space Flight Center, Greenbelt, MD 20771, USA}

%


\begin{abstract}

  We report the discovery of a \period\ pulsar associated with the TeV
  $\gamma$-ray source \tev\ using the {\it Nuclear Spectroscopic
  Telescope Array} (\nustar) X-ray observatory.  \psr\ lies within
  the shell-type supernova remnant (SNR) \snr, and coincides with an
  X-ray point source and putative pulsar wind nebula (PWN) previously
  identified in \xmm\ and \chandra\ images.  It is spinning down
  rapidly with period derivative \pdot, yielding a spin-down
  luminosity \edot, characteristic age \tauc, and surface dipole
  magnetic field strength \bs. For the measured distance of 12~kpc to
  \snr, the $0.2-10$~TeV luminosity of \tev\ is 6\% of the pulsar's
  present $\dot E$.  The \fermi\ source \fhl\ is marginally
  coincident with \psr, but we find no $\gamma$-ray pulsations
  in a search using 5 years of \fermi\
  Large Area Telescope (LAT) data.  The pulsar energetics support
  an evolutionary PWN model for the broad-band spectrum of \tev,
  provided that the pulsar's braking index is $n\approx2$,
  and that its initial spin period was $P_0 \sim 15$~ms.
  \end{abstract}

\section{Introduction}

The detection by the HESS Galactic plane survey \citep{aha05} of
$10^{12}$~eV radiation coming from the diffuse remains of supernovae
has opened up a new window on the evolution of these energetic stellar
remnants.  More than 2/3 of the $>80$ Galactic TeV
sources\footnote{TeVCat, http://tevcat.uchicago.edu/} are supernova 
remnants (SNRs) or pulsar wind nebulae (PWNe), the latter being the most 
numerous class.

High-energy radiation from PWNe is produced as a result of the
interaction of the pulsar wind with the surrounding medium.  Many
young PWNe are found inside shell SNRs, where the emission begins
at a termination shock situated close
to the pulsar.  Downstream of the shock the relativistic electrons
radiate synchrotron photons from radio through X-rays.  The
same electrons up-scatter ambient photons, creating a second, broad
spectral peak in the $\gamma$-ray band.  Some SNR shells also emit
high-energy $\gamma$-rays, although the dominant mechanism there is less
clear.  One possibility is the hadronic scenario, decay of
$\pi^0$ mesons created in collisions between shock
accelerated cosmic-ray protons and thermal gas.  However, leptonic
models involving inverse Compton scattering of ambient photons are
also invoked \cite[for a review see][]{rey08}.


\tev\ \citep{aha06} is coincident with \snr, a
shell-type, $8^{\prime}$ diameter SNR \citep{sha70,whi96}.
The TeV emission was described as resolved, but centrally peaked.
Deeper observations \citep{hess14} show a more extended
source than first reported,
with Gaussian $\sigma$ = $4.\!^{\prime}3\pm0.\!^{\prime}2$
and a flux above 200~Gev of $1.65\times 10^{-11}$~erg~cm$^{-2}$~s$^{-1}$.
Based on 21~cm \ion{H}{1} absorption spectra to \snr\ and adjacent \ion{H}{2}
regions, \citet{lem09} concluded that the distance to \snr\ is in
the range $12-13.5$~kpc, which makes \tev\
the most luminous TeV source in the Galaxy, with
$L(0.2-10\ {\rm TeV})=2.8\times10^{35}\,(d/12\ {\rm kpc})^2$~erg~s$^{-1}$.

An \xmm\ observation by \citet{fun07} identified a highly absorbed
X-ray point source coincident with the \hess\ source with clear
indication of extended emission.  The X-ray components were
subsequently well resolved by \chandra\ \citep{lem09}.
No X-ray emission is detected from the SNR shell, probably because of
its low temperature and large intervening column density, $N_{\rm H}
\sim 1.4\times10^{23}$~cm$^{-2}$. The power-law spectrum of the
\chandra\ point source, with photon index $\Gamma_{\rm
  PSR}\approx1.1$, is consistent with a pulsar origin.  The compact
X-ray PWN is $\approx1.\!^{\prime}2$ in diameter, smaller than the TeV
source, while its spectral index, $\Gamma_{\rm PWN}\approx2.5$, is
much steeper than that of the pulsar.  Its spectrum steepens further
in the outer parts, which is evidence of synchrotron aging.  No radio
counterpart to the X-ray PWN is detected in high-resolution images of
\snr\ using the GMRT at 235, 610, or 1280~MHz, and ATCA maps at 1290
and 2300~MHz, and no radio pulsations were found by the GMRT at 610
and 1280~MHz \citep{cas11}.

The \fermi\ source \fglone, coincident in position with \tev, was
interpreted by \citet{sla10} as leptonic emission from a PWN.  Their
derived GeV spectrum is not cut off as would be the case for
magnetospheric pulsar emission, but is continuous with the HESS 
spectrum. In contrast, \citet{hess14} interpreted the joint \fermi/HESS
spectrum and revised TeV extent in terms of hadronic emission from
a portion of the SNR shell interacting with dense interstellar gas.
More recent analyses of additional \fermi\ data \citep{nol12,ace13,ack13}
show less flux than \citet{sla10} found, and the revision
will affect both models.

In this paper, we present the discovery of X-ray pulsations from
\tev/\snr\ in data acquired as part of the \nustar\ survey of the
Norma Arm region of the Galactic plane \citep{har13,for14}, and a
follow-up observation to determine the pulsar's energetics.  Section~2
describes the \nustar\ observations, pulsar analysis, and spectrum.
In Section~3 we report on an unsuccessful search of the \fermi\ LAT
data for $\gamma$-ray pulsations.  In Section~4 we discuss the
properties of \tev/\snr\ in the context of previous leptonic PWN
models for its broad-band spectrum, and use the measured spin
parameters of \psr\ in a revised evolutionary PWN model.  Some
of the implications of our model, and a comparison of its
assumptions and results with the hadronic model of \citet{hess14},
are presented in Section~5.

\section{\nustar\ Observations}

The field containing \tev\ was observed by \nustar\ in three offset
pointings of $\approx25$~ks each on 2013 June 20--24 as part of a
survey of the Norma region of the Galactic plane.  However, the source
fell in the chip gap for the first observation (ObsID 40014011002)
so we exclude this data set in the final timing and spectral
analysis. 
%
%
A follow-up, dedicated 90~ks observation of \tev\ was performed on
2013 September~29.
An observation log is presented in Table~\ref{tab:obslog}. Data were
collected using \nustar's two co-aligned X-ray telescopes, with
corresponding focal plane modules FPMA and FPMB.  These telescopes
provide $18^{\prime\prime}$ FWHM imaging resolution over the 3--79~keV
X-ray band, with a characteristic spectral resolution of 400~eV FWHM
at 10~keV \citep{har13}.  The reconstructed \nustar\ coordinates are
accurate to $7.\!^{\prime\prime}5$ at 90\% confidence.  The relative
timing accuracy of \nustar\ is limited to $\approx 2$~ms rms, after
correcting for thermal drift of the on-board clock, with the absolute
time scale shown to be better than $<3$~ms \citep{mor14}.


\begin{figure}[t]
\centerline{
\hfill
\psfig{figure=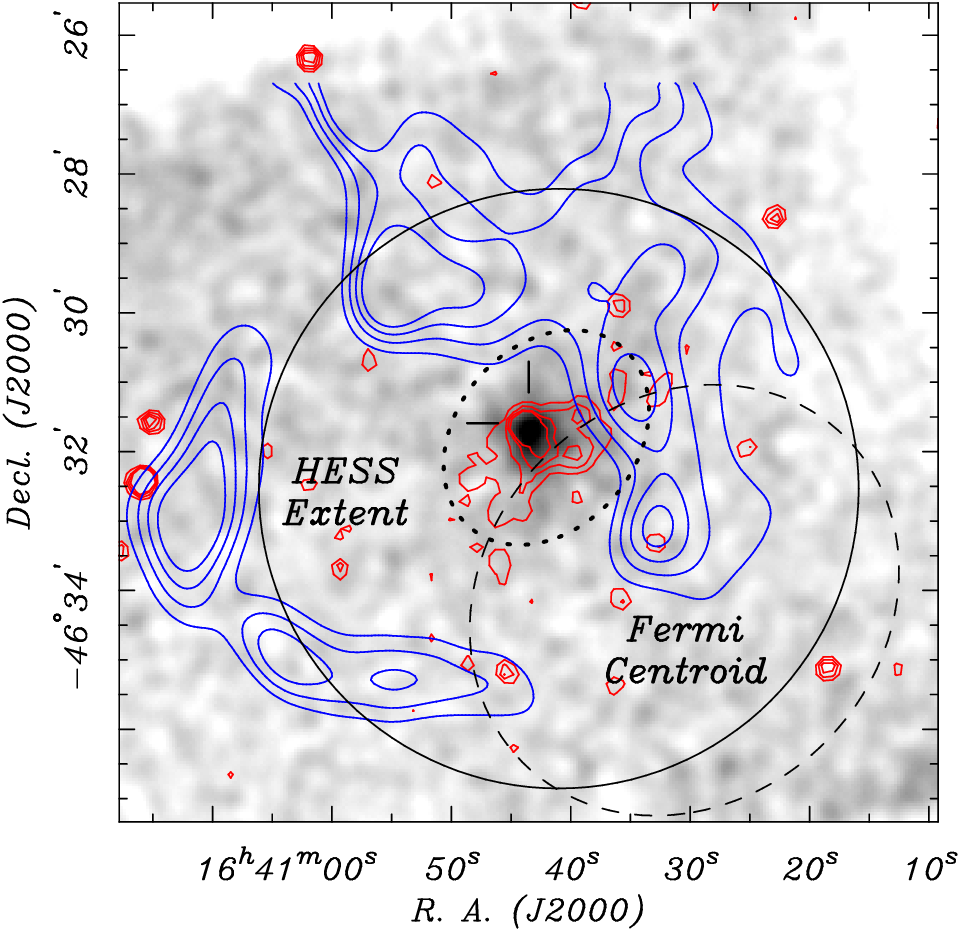,height=0.96\linewidth,angle=-90}
\hfill
}
\caption{\nustar\ 3--79~keV exposure-corrected image of the field
  containing \tev\ acquired on 2013 June 22.  The newly detected
  pulsar is coincident with \cxo\ (tick marks).  Significant
  $0.5-10$~keV emission from the PWN surrounding the pulsar is
  outlined by the \chandra\ contours (red) and enclosed by the
  spectral extraction region (dotted ellipse). SNR \snr\ is indicated
  by the MOST 843 MHz contours (blue, \citealt{whi96}).  The TeV
  extent of \tev\ is denoted by the solid circle \citep{hess14}.
  \psr\ is just outside the 95\% confidence error ellipse of the
  $>10$~GeV \fermi\ source \fhl\ (dashed circle, \citealt{ack13}).}
\label{fig:images}
\end{figure}

Data were processed and analyzed using {\tt NuSTARDAS} v1.3.1 and {\tt
  HEASOFT} v6.15.1 and the Calibration Database (CALDB) files from
2013 August 30. Our analysis used the standard level-1 filtered event
files generated by {\ttfamily nupipeline}.  The observations were free
of significant time variable particle background contamination.
However, each data set has a unique spatial background pattern across
the field-of-view, which affects the sensitivity to a pulsed signal.

%
 
Figure~\ref{fig:images} displays the \nustar\ 3--79~keV image of \tev,
obtained on 2013 June 22 with the source on-axis (ObsID 40014016001).
The image has been corrected for exposure, smoothed using a
$\sigma=7.\!^{\prime\prime}4$ Gaussian kernel, and scaled linearly.
There is one significant source whose position is $7^{\prime\prime}$
from the putative pulsar \cxo, which is coincident within the
\nustar\ positional uncertainty.  There is
evidently diffuse emission surrounding the point source, consistent
with that found in \xmm\ and \chandra\
images, and interpreted as a PWN \citep{fun07,lem09}.  This, plus its
X-ray flux and spectrum (see below) leave no doubt that the \nustar\
source is the counterpart of \cxo.  The most recent \chandra\ image
obtained on 2011 June~6 \citep{for14} shows that this source has not
varied compared to the two previous \chandra\ observations of 2010
June~19 and 2007 May~11.


\begin{deluxetable*}{lcccclc}
\tabletypesize{\small}
\tablewidth{0pt}
\tablecaption{Log of \nustar\ Observations and Period Measurements}
\tablehead{
\colhead{ObsID} &\colhead{Start Date} & \colhead{Exposure/} & \colhead{Rate\tablenotemark{a}} & \colhead{Start Epoch} & \colhead{Period\tablenotemark{b}} & \colhead{$Z^2_1$} \\
 & \colhead{(UT)} & \colhead{Span (ks)} & \colhead{(s$^{-1}$)}  & \colhead{(MJD)}  & \colhead{(s)}  & \colhead{}
}
\startdata
40014016001/17001 & 2013 Jun 22 & 48.6/96.5  & 0.011 & \epochone\  & \perone\  & \ztestone\  \\
30002021002/21003 & 2013 Sep 29 & 89.9/166.4 & 0.011 & \epochtwo\  & \pertwo\  & \ztesttwo\ 
\enddata
\tablenotetext{a}{Background subtracted $3-25$~keV count rate in
a $30^{\prime\prime}$ radius aperture.}
\tablenotetext{b}{Period derived from a $Z^2_1$ test. The Monte-Carlo derived
$1\sigma$ uncertainty on the last digits is in parentheses.}
\label{tab:obslog}
\end{deluxetable*}



\begin{deluxetable}{ll}
\tablecaption{Timing Parameters of \psr }
\tablehead{
\colhead{Parameter}   &
\colhead{Value}
}
\startdata                                       
R.A. (J2000.0)\tablenotemark{a}   & $16^{\rm h}40^{\rm m}43.\!^{\rm s}52$ \\
Decl. (J2000.0)\tablenotemark{a}  & $-46^{\circ}31^{\prime}35.\!^{\prime\prime}4$ \\
Epoch (MJD TDB)                                  & \epochzero\  \\
Frequency\tablenotemark{b}, $f$                  & \freqzero\  \\
Frequency derivative\tablenotemark{b}, $\dot f$  & \fdotv\ \\
Period\tablenotemark{b}, $P$                     & \perzero\  \\
Period derivative\tablenotemark{b}, $\dot P$     & \pdotv\ \\
Spin-down luminosity, $\dot E$                   & \edotv\           \\
Characteristic age, $\tau_c$                     & \taucv\           \\
Surface dipole magnetic field, $B_s$             & \bsv
\enddata
\tablenotetext{a}{\chandra\ position from \citet{lem09}.}
\tablenotetext{b}{$1\sigma$ uncertainties given
in parentheses.}
\label{tab:ephem}
\end{deluxetable}

\begin{figure*}[t]
\centerline{
\hfill
\psfig{figure=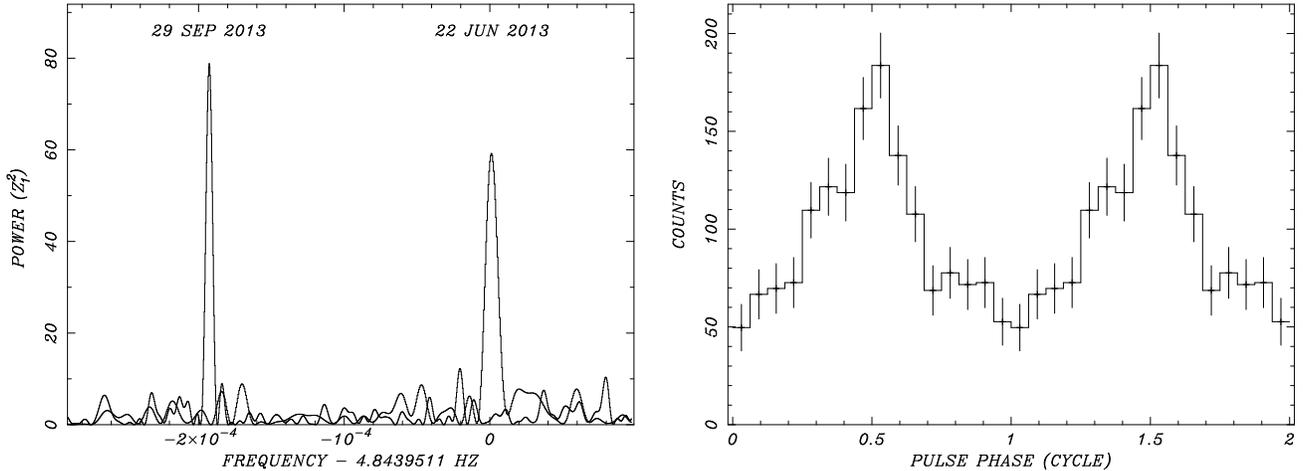,height=0.46\linewidth,angle=270}
\hfill
\psfig{figure=hess1640_nustar_sumfold16.ps,height=0.46\linewidth,angle=270}
\hfill
}
\caption{\nustar\ discovery of \psr.  { Left:} Superimposed power
  spectra for the 2013 June 22 discovery observation and the 2013
  September 29 follow-up showing the significant 4.84~Hz signal and
  its change in frequency. {Right:} The sum of
  background-subtracted pulse profiles from both epochs, each folded
  using the timing parameters in Table~\ref{tab:ephem}, and
  artificially aligned in phase. Phase zero is arbitrary and two
  cycles are shown for clarity.
  Data in both panels include $3-25$~keV photons extracted from a
  $30^{\prime\prime}$ radius aperture.}
\label{fig:timing}
\end{figure*}

\subsection{Timing Analysis}

To search for pulsations from \cxo, we examined \nustar\ data obtained
during the two Norma Arm survey pointings, adjacent in time. The day-long
data set yields a total of 913 photons from the two focal plane
modules, extracted in the optimal $3-25$~keV energy bandpass using a
$30^{\prime\prime}$ radius aperture centered on the source location.
Photon arrival times were corrected for clock drift and converted to
barycentric dynamical time (TDB) using the JPL DE200 ephemeris and the
\chandra\ coordinates of the point source.
The arrival times were binned at 2~ms and searched for coherent
pulsations up to the Nyquist frequency using a $2^{26}$ bin fast
Fourier transform (FFT).  We found a \period\ coherent signal with a
power of $P_{\rm FFT} = 58$, significant at the $99.997\%$
($\simeq4\sigma$) confidence level.  This motivated a second-epoch
observation from which we were able to confirm the detection
($P_{\rm FFT} = 77$) and measure the spin-down rate of the pulsar as
described next.

We computed the spin-down rate from the difference in frequency over
the 100~day time span between the June and September observations. For each
data set, we generated a refined frequency measurement by oversampling
the signal using the $Z^2_1$ test statistic around the known frequency
(Table~\ref{tab:obslog}).
We then re-fitted for frequency by including the frequency derivative
in the light-curve folds, as each observation is sufficiently long
that spin-down smears the pulsed signal ($\Delta \phi = 0.11$ and
0.32~cycles for the first and second epoch, respectively). The final
period measurement used the iterated period derivative reported in
Table~\ref{tab:ephem}.  The uncertainties are estimated from a Monte
Carlo simulation of the light curve using the method described by
\citet{got99}.  The final power spectra are shown in
Figure~\ref{fig:timing} (left).  The derived physical parameters in
Table~\ref{tab:ephem} are from the relations $\dot E=4\pi^2I\dot P
P^{-3}$, where $I=10^{45}$ g~cm$^2$, and the dipole spin-down relations
$\tau_c\equiv P/2\dot P$ and $B_s=3.2\times10^{19}(P\dot P)^{1/2}$~G.

Figure~\ref{fig:timing} (right) shows the highly modulated pulse
profile characterized by a relatively sharp peak and broad trough
compared to a pure sinusoid signal. The pulsed fraction
in the $3-25$~keV band is $f_p>48\pm10\%$ after allowing for the
background level in the source aperture, estimated using data from a
concentric annulus. This increases to $f_p \approx 82\%$ after taking
into account contamination from the PWN in the source aperture using
the spectral results presented below. Here, we define the pulsed fraction
as the ratio of the pulsed emission to the net source (background
subtracted) flux. To determine the unpulsed level we take the average
of the two lowest bins in the 16-bin folded light curve of
Figure~\ref{fig:timing}.  The pulse shape shows no variation
with energy, within statistics.



\newpage

\subsection{Spectral analysis}

For spectral study we extracted photons from the observations listed
in Table~\ref{tab:obslog} using an elliptical region of diameter
$3\farcm4 \times 2\farcm6$ whose major-axis is oriented at position
angle $-40^{\circ}$, centered on the apparent enhanced PWN emission at
(J2000) $16^{\rm h}40^{\rm m}42.\!^{\rm s}05,
-46^{\circ}31^{\prime}47.\!^{\prime\prime}8$, which is offset by
$20^{\prime\prime}$ from the pulsar (see Figure~\ref{fig:images}).  We
estimate the background using nearby $50^{\prime\prime}$ radius
circular regions carefully chosen to account for stray light in the
focal planes in two of the observations.  Response matrices were
generated for each spectral file using the \nustar\ analysis software.
All eight spectra were combined, as were their matching response
matrices, and fitted using the XSPEC software package \citep{arn96}
with $\chi^{2}$ minimization. The fitting is limited to the 3--20~keV
range due to low signal-to-noise at higher energies. In this range,
the combined spectrum yields a total of 14,100 source plus background
photons, and is grouped to obtain a minimum  detection
significance of $5\sigma$ per spectral bin.

The \nustar\ spectrum in the large aperture is dominated by PWN
emission and is well fit by an absorbed power-law model as expected
for non-thermal emission from the nebula. We used the {\tt tbabs}
absorption model, with \cite{wam00} abundances and \cite{vern96} cross
sections, and obtain a best-fit $N_{\rm
  H}=(1.7\pm0.9)\times10^{23}$\,cm$^{-2}$ and $\Gamma=1.9\pm0.4$.
Errors are 90\% confidence level
($\Delta\chi^{2}=4.61$
for two interesting parameters) determined from the error ellipse contour
extrema. This provides an acceptable fit with a reduced
$\chi^{2}_{\nu}=0.87$ for 36 degrees of freedom (dof). The absorbed
2--10\,keV flux is $(8.0^{+0.4}_{-2.0})\times10^{-13}$\,erg\,cm$^{-2}$\,s$^{-1}$
(90\% confidence).
This is consistent with the values reported by \cite{fun07} using
\xmm\ data extracted from a $2\farcm5$ diameter aperture centered on
the PWN.

To further constrain the PWN emission and to isolation the pulsar
contribution to the \nustar\ spectrum, we include the \chandra\
spectrum of the pulsar and PWN in the fitting process. We supplemented
the \chandra\ data set (ObsID~7591) previously analyzed as part of the
study of \tev\ by \cite{lem09} with data obtained on 2011 June 6
(ObsID~12508).  These data were acquired with the Advanced CCD Imaging
Spectrometer \citep[ACIS,][]{garmire03} and reprocessed using the
{\ttfamily chandra\_repro} package of the \chandra\ Interactive
Analysis of Observations (CIAO) software suite.  Spectra and their
response matrices from the two observations were produced using
{\ttfamily specextract}. The pulsar and PWN spectra are grouped with a
minimum of 10 and 40 counts per spectral bin, respectively, and each
fitted with the absorbed power-law model, using a common column density.

We used an extraction radius of $2^{\prime\prime}$ for the pulsar; the
background is negligible in this small region. For the PWN we used the
\nustar\ elliptical region and similar background region. We combined
spectra from the individual observations, and their response matrices,
as above.  Although the ACIS-I detector is sensitive in the
0.5--10~keV energy range, we restricted the spectral fits to the
3--7~keV range due to the high absorption and limited statistics. A
total of 142 and 1369 photons were fitted for the pulsar and PWN,
respectively, during the 47~ks exposure.  The second column of
Table~\ref{tab:specfit} presents the resulting spectral fits for the
pulsar and PWN spectra using the \chandra\ data alone\footnote{The
  fluxes described in \cite{lem09} as unabsorbed are not consistent with
  the (absorbed) fluxes given in Table~\ref{tab:specfit} or with those
  presented by \cite{fun07}; the \cite{lem09} values are likely
  absorbed fluxes, mischaracterized.}.

\begin{figure}[t]
\centerline{
\hfill
\psfig{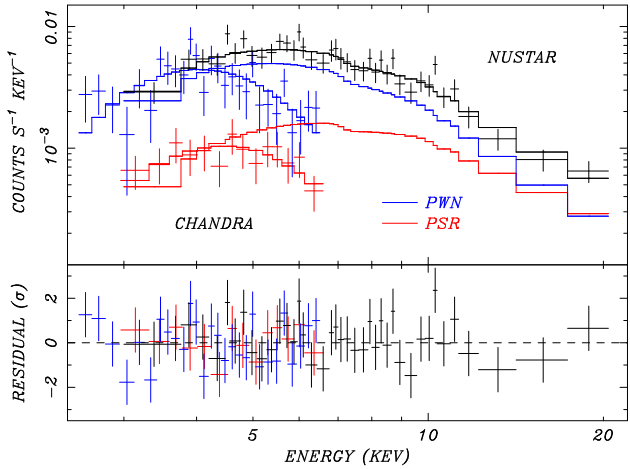}
\hfill
}

\caption{\chandra\ and \nustar\ spectra of \psr\ (red) and its wind
nebula (blue) fitted simultaneously with the absorbed power-law models
presented in Table~\ref{tab:specfit}. The pulsar and PWN
components of the \nustar\ spectrum sum to the total \nustar\ spectrum (black).
The \chandra\ and \nustar\ fits are tied in
the 2--10~keV band, but allow for a constant flux offset
between missions. The lower panel shows residuals from the best fit.
}
\label{fig:spectra}
\end{figure}

We fitted the \chandra\ and \nustar\ spectra together using two
power laws as before, with their column densities tied. To constrain the
pulsar and PWN contributions to the \nustar\ spectrum, we fixed the
power-law indices to the \chandra\ models and tied their flux together
in the overlapping 2--10~keV band. To allow for flux calibration
differences between the two missions we introduce an overall
normalization constant to the \nustar\ model, with best fit value of
$1.11$. The resulting spectrum is shown in Figure~\ref{fig:spectra},
and the parameters are reported in Table~\ref{tab:specfit}. The
\nustar\ spectra of the pulsar and PWN are evidently successfully
modeled, as the fitted values are in good agreement with the \chandra\
results for each component. To estimate the PWN contribution to the pulse
profile shown in Figure~\ref{fig:timing}, we repeated our spectral
analysis using a $30^{\prime}$ radius aperture; the PWN is found to
account for $\approx53\%$ of the background-subtracted source flux in
pulse profile.

\begin{deluxetable}{lcc}
\tablecaption{X-ray Spectrum of \psr\ and its Wind Nebula.} 
\tablecolumns{3}
\tablewidth{0.96\linewidth}
\tablehead{ \colhead{Parameter}   &  \colhead{\chandra\ only}  &  \colhead{\chandra\ +\nustar} }
\startdata
$N_{\rm H}$ (cm$^{-2}$)                       & $(1.2\pm 0.6) \times 10^{23}$         & $(1.8\pm 0.6) \times 10^{23}$        \\
$\Gamma_{\rm PSR}$                            & $1.2^{+0.9}_{-0.8}$                   & $1.3^{+0.9}_{-0.5}$                  \\
$F_{\rm PSR}$ (2--10 keV)                     & $1.9^{+0.2}_{-1.4} \times 10^{-13}$        & $(1.8\pm 0.4) \times 10^{-13}$       \\
$\Gamma_{\rm PWN}$                            & $2.3^{+1.2}_{-1.0}$                   & $2.2^{+0.7}_{-0.4}$                  \\
$F_{\rm PWN}$ (2--10 keV)                     & $5.4^{+0.6}_{-2.3} \times 10^{-13}$& $(5.5\pm 0.8) \times 10^{-13}$       \\
$\chi^2_{\nu}$ (dof)                          & 1.0 (56)                              & 0.82 (79)                                
\enddata
\tablecomments{Absorbed power-law model fits to the pulsar and the PWN
  spectra with their column densities linked. The simultaneous fit to
  \nustar\ and \chandra\ data is described in the text. The
  uncertainties are 90\% confidence limits determined from the error
  ellipse extrema.  The given fluxes are absorbed, in units of
  erg~cm$^{-2}$~s$^{-1}$. }
\label{tab:specfit}
\end{deluxetable}

\section{Search for Gamma-ray Pulsations}


Although \cite{sla10} argue that the GeV emission from \fglone\
originates from the PWN, the \fermi\ source is spatially unresolved
and marginally consistent with the position of the pulsar.  Therefore,
we searched the \fermi\ LAT data at the position of \psr\ for a pulsed
signal. The X-ray timing parameters and their errors given in
Table~\ref{tab:ephem} allow a search for pulsations around the known
values. We extract photon arrival times from 2008 August~4 to 2013
October~2.  These data were reprocessed using the ``Pass~7'' event
reconstruction algorithm. We selected ``source'' photons for zenith
angles $<100^{\circ}$ and restricted the energy range to $>500$~MeV to
minimize Earth limb and diffuse
Galactic $\gamma$-ray background contamination. The photon arrival
times were corrected to the solar system barycenter using the JPL
DE405 ephemeris and the \chandra\ coordinates. For our nominal pulsar
search we extracted photons from an energy-dependent radius enclosing
95\% of the point spread function.

Because the sparse X-ray observations do not provide a phase-connected
ephemeris, unknown timing noise could limit the practical time span of
a coherent pulsar search. Accordingly, we grouped the photons into
intervals of 100 days and searched for a significant signal over a
range of $f,\dot f$ centered on interpolated values at the test epoch
using the X-ray timing parameters.  This range corresponds to three
times the $1\sigma$ uncertainty on each of $f$ and $\dot f$.  We used
the $H$-test statistic \citep{dej10} to allow for a complex
profile with narrow features. This test selects the Fourier harmonic
$m \leq 20$ that results in the most significant normalized $Z^2_m$ power.

None of the resulting searches yielded a significant signal above the
noise.  We also summed the power from these searches for each $f,\dot
f$ pair, to increase the signal sensitivity.  No significant signal
was apparent.  To explore a range of instrumental parameters that
might be more sensitive, we repeated our search for a low-energy
cutoff of 100, 300, and 500 MeV, and time interval of 30, 200, and
300 days. Despite the expanded range of parameter space, no signal
stronger than the expected noise is detected in each search or in the
summed results.

\section{Modeling the Broad-band Spectrum}

\subsection{Age, Energetics, and Distance}

The discovery of \psr\ supports the conjecture that \cxo, \tev, and
\fhl\ are all manifestations of a middle-aged pulsar/PWN system. For
the measured distance of 12~kpc to \snr\ (see below), the $0.2-10$~TeV
luminosity of \tev\ is 6\% of the pulsar's present $\dot E$, while the
ratio $F(0.2-10\,{\rm TeV})/F_{PWN}(2-10\,{\rm keV})\approx13$,
indicating that, in a leptonic model where the TeV emission results
from up scattering of ambient photons by relativistic electrons,
inverse Compton losses now dominate over synchrotron emission.

A result that was perhaps unexpected is the young characteristic age
of the pulsar.  The pulsar's characteristic age, \tauc, is an
approximate measure that applies when a simple vacuum dipole spin-down
model is assumed, and when the spin period is significantly larger
than that at birth.  Although pulsars for which the model has been
tested do not rotate like simple vacuum dipoles \citep{kas01},
it is notable that, excluding magnetars, there are only seven pulsars
with $\tau_c<3000$~yr in the Australia Telescope National Facility (ATNF)
catalog\footnote{http://www.atnf.csiro.au/research/pulsar/psrcat/expert.html}
\citep{man05},
followed by \psr\ with $\tau_c=$~\taucv.  This is younger than most
authors have assumed in modeling the evolution of \tev/\snr\
($T=20$~kyr, \citealt{fun07}; $T=15$~kyr, \citealt{lem09};
$T=10$~kyr, \citealt{sla10}).
\citet{fan10} presented two models similar to that of \citet{sla10},
one with $\dot E=1\times10^{38}$~erg~s$^{-1}$ and $T=4500$~yr, and the
other with $\dot E=1.65\times10^{37}$~erg~s$^{-1}$ and $T=8200$~yr.
While these ages are closer to the likely true age, the actual
spin-down power of \psr\ is considerably smaller than the assumed
values.

Prior models for \snr\ estimate the age of the SNR from
its radius assuming that it is in the Sedov phase with
$r_s=(2.02\,{\cal E}/\rho)^{1/5}\,t^{2/5}$ \citep{spi78}.
These estimates are imprecise due to the
unknown energy $\cal E$ of the explosion and the ambient
mass density $\rho$, and the strong dependence of age $t$ on radius, 
which magnifies any uncertainty in distance.
Also, the pulsar is offset from the center of the SNR in a direction
that, given the morphology of the PWN, cannot be explained by high
kick velocity.  \citet{lem09} conclude that the geometrical center
of the remnant is not the explosion site.  Instead, the structure
of the remnant may be affected by complicated interactions
with the local ISM, which could affect the estimation of its age.

There is no near/far distance ambiguity for \snr\
because 21~cm absorption is seen up to the maximum negative
value at the tangent point in both it and adjacent \ion{H}{2} regions.
\citet{lem09} concluded that the distance to \snr\ is in
the range $12-13.5$~kpc, while \citet{kot07} had already found a
value of $\approx11-12$~kpc from the same 21~cm data, but assuming a
Galactic center distance of 7.6~kpc instead of 8.5~kpc.
We adopt $d=12$~kpc here.
While the distance appears to be constrained well, there is
no direct measurement of the ambient density; the above referenced models
assume values of $n_{\rm ISM}$ between 0.1~cm$^{-3}$ and 10~cm$^{-3}$.
\citet{cas11} estimated an electron density of 100--165~cm$^{-3}$
for the western and northern part of the SNR shell by assuming
that the low-energy turnover of its radio spectrum is due to local
free-free absorption.  They also searched the NANTEN CO survey
data for molecular gas that could be
target material for hadronic production of $\gamma$-rays
from protons accelerated in the SNR shell,
but did not find any that could be associated with \snr.

\subsection{A Hadronic Model}

\citet{hess14} present a model in which the TeV emission is produced by
hadronic interactions in the north and west part of the SNR shell.
It is motivated largely by the continuity of the \citet{sla10}
\fermi\ spectrum with the HESS spectrum, which they argue is difficult
to fit in a PWN model, and also by the overlap of the TeV
source with the side of the SNR shell that is adjacent to \ion{H}{2} regions.
They require a high ambient density of 150~cm$^{-3}$ to reproduce the
GeV/TeV spectrum, but a small density to account for the size of
the SNR shell.  These are achieved by assuming that the explosion
occurred inside a wind-blown bubble of the progenitor star,
of average density 0.1~cm$^{-3}$,
which is then consistent with their assumed age of 2500~yr.  As they
remark, the model requires a large fraction of the SN energy
to be channeled into cosmic rays, especially since only one
side of the SNR contributes to this process.  However, they
overestimated the integrated $>1$~TeV flux of \hess\ by a factor
of $\approx6$ due to a numerical error, which relieves
the efficiency requirement somewhat.  We calculate a 
 $>1$~TeV luminosity of $1.2\times10^{35}\,(d/12\ {\rm kpc})^2$~erg~s$^{-1}$
instead of their $4.6\times10^{35}\,(d/10\ {\rm kpc})^2$~erg~s$^{-1}$
(note also the different assumed distances).

As we argue below, the \citet{sla10} \fermi\ spectrum used
by \citet{hess14} is not likely to be correct, and there may not
even be a detectable source in \fermi\ at energies $<10$~GeV.  This
is why we use different \fermi\ results in our own model,
which contributes to the diverging conclusions
of the two papers.

\subsection{Leptonic PWN Models}

\citet{sla10} modeled the $\gamma$-ray emission from \hess\ and the
associated broad-band spectrum with an evolving, one-zone PWN model,
where the $\gamma$-rays are ambient photons inverse Compton scattered 
by the relativistic electrons that also produce the
synchrotron emission.  Crucial to these models are the age and
spin-down power of the pulsar.  \citet{sla10} assumed a remnant age of
10~kyr and a pulsar with spin-down power of $\dot
E=4\times10^{36}$~erg~s$^{-1}$.  The latter came from the $\dot E/L_x$
relation of \citet{pos02} and, given the observed scatter, it was
fortuitously a very accurate prediction.  It
was necessary to add a Maxwellian distribution of electrons around
0.1~TeV to a power-law tail to account for the strong \fermi\ GeV
$\gamma$-rays that they found from this source relative to its TeV flux. 

The pulsar age assumed by \citet{sla10} was likely too large.
The true age of a pulsar is given by 
$$T = {P \over (n-1)\dot P}\left[1-\left(P_0 \over P\right)^{n-1}\right],$$ 
where $n\equiv f\ddot f/\dot f^2$ is the braking index, assumed by
Slane et al. (2010) to be 3, consistent with a vacuum dipole, $P_0$
is the spin period at birth, and $P$ is the currently measured period.
For $n=3$ the pulsar's true age is less than its characteristic age,
and refined models will likely favor a younger system and/or a smaller
braking index.


\begin{deluxetable*}{lllc}[t]
\tablecaption{Properties of \tev} 
\tablehead{ \colhead{Parameter}   &  \colhead{Observed} &  \colhead{Modeled}
& \colhead{Reference} }
\startdata
SNR radius      & $4\farcm5 \pm 0\farcm5$ & $4\farcm51$ & \citet{sha70} \\
$S_{\nu}$ [660~MHz] (mJy)        & $\leq 690$ & $337$ & \citet{gia08} \\
$F_X$ [2--25 keV] (erg~cm$^{-2}$~s$^{-1}$)  & $(1.68\pm0.4)\times10^{-12}$ & $1.0\times10^{-12}$ & This work \\
$\Gamma_X$                            & $2.2^{+0.7}_{-0.4}$ &$2.4$ & `` \\
$F_{\gamma}$ [10--500 GeV] (erg~cm$^{-2}$~s$^{-1}$) & $(3.15\pm1.00)\times10^{-11}$ & $1.6\times10^{-11}$  &  \citet{ack13} \\
$\Gamma_{\gamma}$              & $1.92\pm0.24$ & $1.96$ & ``
\enddata
\tablecomments{The quoted 2--10 keV flux is corrected for interstellar absorption.}
\label{tab:modeldata}
\end{deluxetable*}

Assuming again that the PWN of \psr\ is primarily responsible for the
$\gamma$-ray emission, we fit an evolutionary model of a PWN inside
an SNR to the broad-band spectrum. This can constrain the properties
of the central neutron star, the pulsar wind, progenitor supernova,
and the surrounding ISM.  We include the PWN emission in
the X-ray band and the radio upper-limit reported in the literature
(see Table~\ref{tab:modeldata}).  However, in
examining more recent \fermi\ publications,
we note that the source \fgltwo\ \citep{nol12} is fainter
than the flux (based on $\sim 1$~yr of data)
that was extracted by \citet{sla10}.  More
recent analyses of this \fermi\ source
have been restricted to energies $>10$~GeV \citep{ace13,ack13},
and we use only the latter in our revised modeling,
in particular the spectrum of \fhl\ \citep{ack13} derived
from $3$~yr of data.
Indeed, it is not obvious from inspecting \fermi\ counts
maps that there is a significant source at energies
$<10$~GeV, which is why we do not attempt to model this
part of the \fermi\ band.
This is an important difference from \citet{hess14},
who accept and model the \fermi\ results of \citet{sla10}.

Our model is based on the work of \citet{gel09}, modified to
include background photon fields in addition to the cosmic microwave
background (CMB), and a broken power-law spectrum of particles injected
into the PWN at the termination shock, as
favored by recent studies of these objects (e.g.,
\citealt{buc11}).  We assumed two additional background photon
fields: one with temperature $T_1=15~{\rm K}$ and energy density
$u_1=4u_{\rm CMB}$, where $u_{\rm CMB}=4.17\times10^{-13}~{\rm erg\ cm^{-3}}$
is the energy density of the CMB, and the other with
$T_2=5000~{\rm K}$ and $u_2 = 1.15u_{\rm CMB}$, the same as used by
\citet{sla10}.  We fixed the distance at 12~kpc.  To
obtain the best-fit values of the free parameters and their
errors, which are listed in Table~\ref{tab:modelparam},
we employed a Markoff Chain Monte Carlo algorithm similar to the
one used to fit the properties of the PWNe in Kes~75
\citep{gel13} and G54.1+0.3 (Gelfand et al., in preparation).

For a constant braking index, the spin-down power of the
pulsar evolves as
$$\dot E(t)=\dot E_0\left(1+{t \over \tau_0}\right)^{-({n+1 \over n-1})},$$
where $\dot{E}_0$ is the initial spin-down power and $\tau_0$
is the initial spin-down timescale.  The present $P$ and $\dot P$
are known, which fixes the values of $\dot E_0$ and the
present age $T$ corresponding to each trial value of $n$ and $\tau_0$.

\begin{figure}
\centerline{
\hfill
\psfig{figure=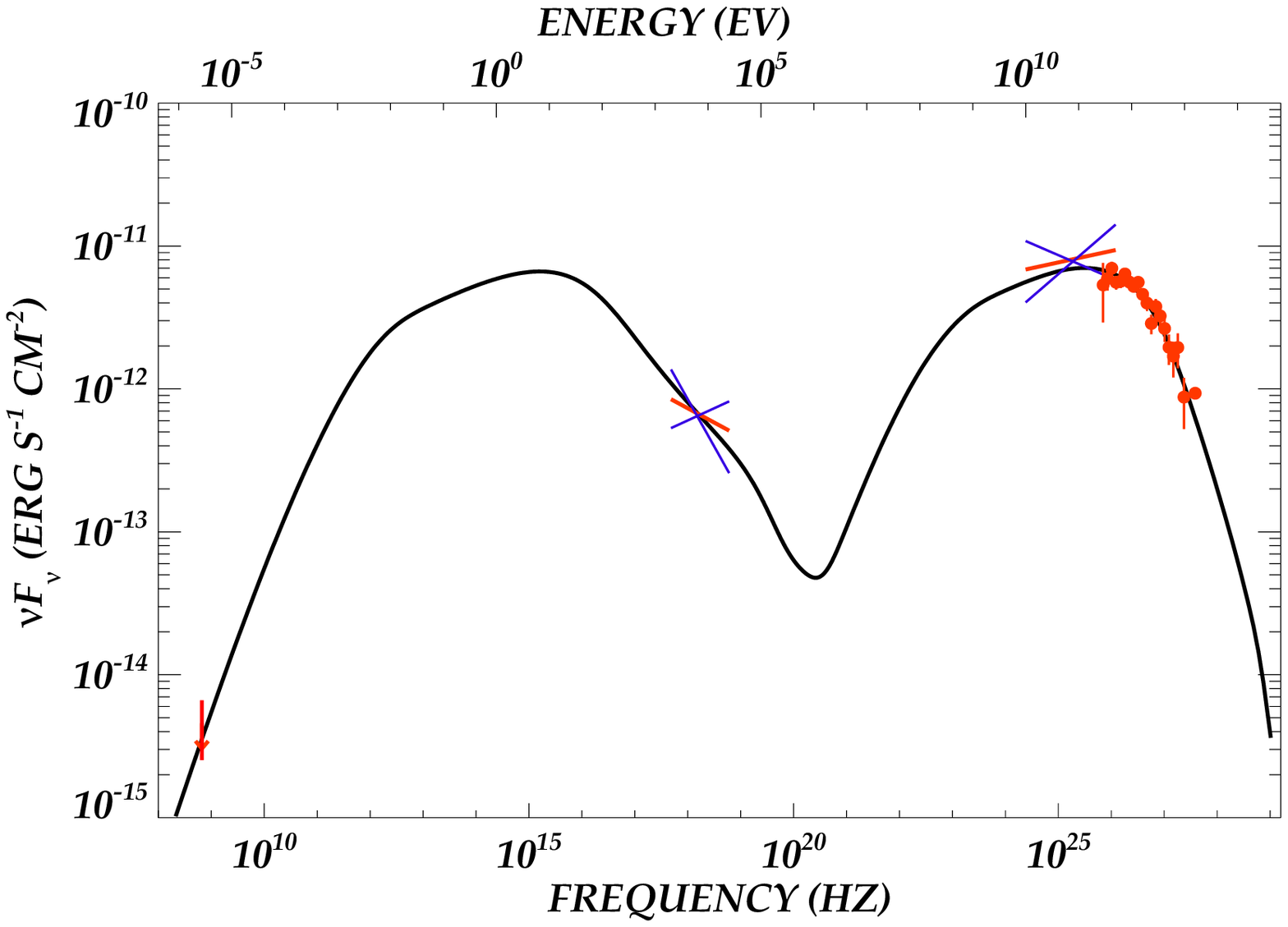,width=0.99\linewidth}
\hfill
}
\caption{Broad-band PWN model fit to the spectrum of \tev. From left to
  right, in red, the radio upper-limit (\citealt{gia08}), the \chandra\
  X-ray spectrum of the PWN (\citealt{lem09}), the \fermi\ GeV spectra
  (\citealt{ack13}), \and the \hess\ TeV data points (\citealt{hess14}).
  The blue lines represent the error range on the \chandra\ and \fermi\
  spectral fits.  The black line is the spectral energy distribution for
  the best-fit model parameters given in Table \ref{tab:modelparam}.}
\label{fig:modelspec}
\end{figure}

As shown in Figure~~\ref{fig:modelspec}, the model is able to
reproduce the observed broad-band spectrum with $\chi^2=24.7$
for 15 degrees of freedom,
using the best-fit parameters given in Table~\ref{tab:modelparam}.
In general, the values of these parameters are similar to what has been
inferred for other systems using similar models (e.g.,
\citealt{buc11}).  However, our best-fit model requires a small
braking index $n\approx1.9$ and a short initial spin-down timescale
$\tau_0\sim800~$~yr.  This implies that the pulsar was
born spinning rapidly, with $P_0\approx15$~ms and
a high initial spin-down power
$\dot{E}_0\approx10^{40}$~erg~s$^{-1}$.  Its present age would
then be 6800~yr.  We note that, in this case, the model is
not entirely self-consistent, as the initial rotational energy
of the pulsar, $9\times10^{49}$~erg, is a significant fraction
of the fitted SN explosion energy, and could affect the
dynamics of the explosion.   


\begin{deluxetable}{lc}
\tablecaption{PWN Evolutionary Model Results for \tev}
\tablehead{ \colhead{Parameter} &  \colhead{Value}  }
\startdata
SN explosion energy, log (${\cal E}/10^{51}$~erg)& $-0.78$ \\ 
Ejecta mass, log ($M_{\rm ej}/M_{\odot}$)        & $-0.34$ \\
ISM density, log ($n_{\rm ISM}$/cm$^{-3}$)       & $-1.53$ \\
Braking index, $n$                               & $+1.87$ \\
Spin-down timescale, log ($\tau_0$/yr)           & $+2.89$ \\
Wind magnetization, log ($\eta_B$)               & $-1.76$ \\
Low-energy particle index, $p_1$                 & $+0.28$ \\
High-energy particle index, $p_2$                & $+2.59$ \\
Minimum energy, log ($E_{\rm min}$/GeV)          & $-3.86$ \\
Break energy, log ($E_{\rm br}$/TeV)             & $-0.71$ \\
Maximum energy, log ($E_{\rm max}$/PeV)          & $+0.12$
\enddata
\label{tab:modelparam}
\end{deluxetable}

\begin{figure}
\centerline{
\hfill
\psfig{figure=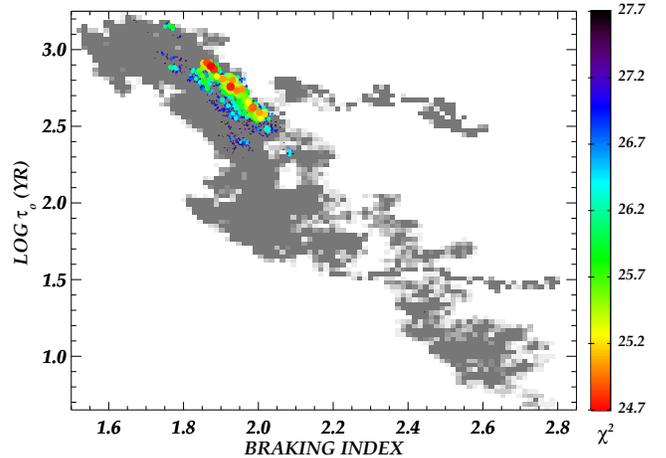,width=0.99\linewidth}
\hfill
}
\caption{Trial values of the braking index $n$ and spin-down timescale
  $\tau_0$ of \psr.  The gray-scale indicates the
  distribution of trials with $\chi^2>27.7$, while the colors
  indicate the $\chi^2$ of the particular trials with $\chi^2\le27.7$.} 
\label{fig:modelp-tau}
\end{figure}


\section{Discussion}

A PWN model for \tev\ is easier to accommodate now because we have
deprecated the original \fermi\ spectrum that was used
in all previous model fitting.  Even though that spectrum appeared
to be continuous with the new HESS points, fitting a single power law,
it is now evident that additional \fermi\ data above 10~GeV are
continuous with the HESS data, but with spectral curvature to lower energy
that is characteristic of lepton cooling. As to the spatial distribution
of TeV photons, while they overlap with part of the SNR shell of \snr,
they also overlap with the X-ray PWN, so it is entirely possible
that both the PWN and the shell contribute to the $\gamma$-ray emission.
Furthermore, it cannot be ruled out that \psr\ emits some pulsed
$\gamma$-rays around 1~GeV that we have not yet been able to
discover.

The modeled SN explosion is energetic enough for
\snr\ to have reached its present radius of $14$~pc in 6800~yr.
The fitted ambient density is $\approx 0.03$~cm$^{-3}$, similar to the value
assumed by \citet{hess14}, so this parameter is not a discriminant
between the models.  The rapid spin-down of \psr\ helps
to explain why its PWN can be such a luminous $\gamma$-ray source.
The electrons currently emitting inverse Compton
scattered TeV photons were injected when the
pulsar's $\dot E$ was much higher,
while the PWN has expanded, reducing its magnetic field strength
and limiting its synchrotron losses.
A similar scenario may explain why the young magnetars
SGR 1806$-$20 \citep{row12} and CXOU J171405.7$-$381031 in the SNR CTB~37B
\citep{hal10} may power TeV sources even though their present spin-down
luminosities are small. 


If $n<3$ for \psr, as it is for
all pulsars in which it has been measured,
one reason could be that its magnetic dipole field strength
or the inclination angle $\alpha$ of its axis is
increasing in time such that
$(B_s\,{\rm sin}\,\alpha)^2\propto P^{3-n}$ \citep{liv07}.
If $n=2$, for example, $\dot P$ will remain constant and
\psr\ will join the group of magnetars when its period is
$\approx7$~s and $B_s=8\times10^{13}$~G.
(However, aside from the idealized $n=3$ case, there is no
reason why $n$ should be constant over the lifetime
of a pulsar.)  The smallest measured pulsar braking index
that is not affected by glitching
is $n=0.9\pm0.2$ for PSR J1734$-$3333 \citep{esp11}.
The 1.17~s period of PSR J1734$-$3333, and its inferred
dipole magnetic field of $\approx 5\times10^{13}$~G,
already close to those of magnetars.

The complex of \ion{H}{2} regions and SNRs including \snr\
is at the intersection of the
far end of the Galactic bar with the Norma spiral arm.
\citet{dav12} detected a young, massive star cluster in the near-infrared
at the center of the \ion{H}{2} region G338.4+0.1, and displaced by 
$8^{\prime}$ from \tev, or a projected separation of 22~pc at the
distance they inferred of $11\pm2$~kpc.  The most massive star in
the cluster is a WR star with an estimated initial mass of
$62\,M_{\odot}$ and an age of 3.7~Myr. The optical extinction
to the cluster is compatible with the X-ray measured column density
to \cxo.  Optical and infrared emission from the \ion{H}{2} regions
and star cluster could provide significant target photons for
inverse Compton scattered TeV emission.  \citet{dav12} suggested
that the progenitor of the putative neutron star could have
been born in, and dynamically ejected from the cluster
during its formation, in which case its mass must have
been at least as large as the most massive star presently in the cluster.
But they stopped short of making this claim because they could not prove
the birth of the pulsar progenitor in the cluster, as opposed to in
a nearby site of independently seeded star formation.

Nevertheless, it is possible to speculate that the large
magnetic field strength of \psr, and the short initial spin period
needed to explain its powerful TeV nebula, may both result
from a massive progenitor such as a WR star.
\citet{dun92} proposed that millisecond initial spin periods
are needed for a dynamo to generate magnetar strength $B$-fields.
There is evidence that the most massive stars
($>40\,M_{\odot}$) are the progenitors of magnetars
\citep{fig05,gae05,mun06}, although there are
exceptions \citep{dav09}.  If \psr\ is found to have
these birth properties, it could be evidence of
a physical link between magnetars and other
high $B$-field pulsars.

\section{Conclusions and Future Work}

\nustar\ is a sensitive instrument for the discovery of
young pulsars in distant parts of the Galaxy that
are obscured by large ISM column densities.
The detection of \psr\ in \snr\ provides long-awaited
evidence that a PWN powers the TeV source \tev,
and its properties test evolutionary models that
fit the multiwavelength spectrum.
The spin-down power of \psr, \edotv, is
typical of the range of middle-aged pulsars powering TeV
nebulae, but its characteristic age of \taucv\ is younger
than the modeled age of the system, which is $\approx 6800$~yr,
with a corresponding braking index of $n\approx2$.
Its observed rapid spin-down and inferred short initial
spin period of $P_0\sim15$~ms make it possible for \psr\ to power the most
luminous known TeV source in the Galaxy.  Nevertheless,
a more sensitive determination of the $\gamma$-ray spectrum below 10~GeV,
and a map of
the spatial distribution of the TeV photons, are needed to discriminate
among models that locate the emission in the PWN versus the SNR shell.
It is possible that two (or more) mechanisms may contribute to the
extraordinary $\gamma$-ray luminosity of \tev.

If radio pulsations are detected and monitored, or with the dedicated
program of timing \psr\ with \nustar\ in progress, we can hope to
measure its braking index, and thereby better estimate the actual age
of \snr\ and the initial spin period of the pulsar.  This can provide
confirmation of the model presented in this work.

\acknowledgements This work made use of data from the \nustar\
mission, a project led by the California Institute of Technology,
managed by the Jet Propulsion Laboratory, and funded by the National
Aeronautics and Space Administration. This work has also made use of
archival data from the \chandra\ X-ray Observatory. Partial support
for E.V.G. and J.A.T were provided by NASA through \chandra\ Award
Numbers SAO GO2-13097X and GO1-12068A, respectively, issued by the
\chandra\ X-ray Observatory Center, which is operated by the
Smithsonian Astrophysical Observatory for and on behalf of NASA under
contract NAS8-03060.  E.V.G acknowledges support by \fermi\ Guest
Investigator Grant NNX12AO89G. Radio contours were obtained from
Molonglo Observatory Synthesis Telescope (MOST) data provided by the
University of Sydney with support from the Australian Research Council
and the Science Foundation for Physics within The University of
Sydney.

\end{document}